\documentclass[apjl,iop]{emulateapj-rtx4}
\usepackage{apjfonts}
\usepackage{lscape}

\usepackage{amsmath,amssymb}
\usepackage{xspace}

\newcommand{\hbsp}{H$\beta$\xspace}
\newcommand{\hg}{H$\gamma$}

\newcommand{\hd}{H$\delta$}

\newcommand{\heps}{H$\epsilon$}

\newcommand{\hzeta}{H$\zeta$}

\newcommand{\htheta}{H$\theta$}

\newcommand{\kms}{\ensuremath{\,\text{km}\ \text{s}^{-1}}}
\newcommand{\obj}{\#254025\xspace}
\newcommand{\cahk}{\ensuremath{\text{\ion{Ca}{2}}\;\text{{H+K}}}\xspace}
\newcommand{\cah}{\ensuremath{\text{\ion{Ca}{2}}\;\text{{H}}}\xspace}
\newcommand{\cak}{\ensuremath{\text{\ion{Ca}{2}}\;\text{{K}}}\xspace}

\newcommand{\re}{\ensuremath{r_\text{e}}}

\newcommand{\hst}{\textit{HST}}

\defcitealias{bruzual:2003}{BC03}
\defcitealias{maraston:2005}{M05}

\slugcomment{To Appear in ApJ Letters}

\shorttitle{A $z=1.82$ Analog of Local Ultra-massive Elliptical Galaxies}
\shortauthors{Onodera et al.}

\begin{document}

\title{
  A $z=1.82$ Analog of Local Ultra-massive Elliptical Galaxies\,\altaffilmark{1}
}

\author{
  M.~Onodera\altaffilmark{2}, 
  E.~Daddi\altaffilmark{2}, 
  R.~Gobat\altaffilmark{2}, 
  M.~Cappellari\altaffilmark{3}, 
  N.~Arimoto\altaffilmark{4,5}, 
  A.~Renzini\altaffilmark{6}, 
  Y.~Yamada\altaffilmark{4}, 
  H.~J.~McCracken\altaffilmark{7}, 
  C.~Mancini\altaffilmark{6}, 
  P.~Capak\altaffilmark{8}, 
  M.~Carollo\altaffilmark{9}, 
  A.~Cimatti\altaffilmark{10}, 
  M.~Giavalisco\altaffilmark{11}, 
  O.~Ilbert\altaffilmark{12}, 
  X.~Kong\altaffilmark{13}, 
  S.~Lilly\altaffilmark{9}, 
  K.~Motohara\altaffilmark{14}, 
  K.~Ohta\altaffilmark{15}, 
  D.~B.~Sanders\altaffilmark{16}, 
  N.~Scoville\altaffilmark{17}, 
  N.~Tamura\altaffilmark{18},
  and
  Y.~Taniguchi\altaffilmark{19}
}

\altaffiltext{1}{
  Based on data collected at Subaru Telescope, 
  which is operated by the National Astronomical Observatory of Japan (S09A-043).
}

\affil{$^2$CEA, Laboratoire AIM-CNRS-Universit\'e Paris Diderot, Irfu/SAp, Orme des Merisiers, F-91191 Gif-sur-Yvette, France}
\affil{$^3$Sub-Department of Astrophysics, University of Oxford, Denys Wilkinson Building, Keble Road, Oxford OX1 3RH, United Kingdom}
\affil{$^4$National Astronomical Observatory of Japan, Osawa 2-21-1, Mitaka, Tokyo, Japan}
\affil{$^5$Graduate University for Advanced Studies, Osawa 2-21-1, Mitaka, Tokyo, Japan}
\affil{$^6$INAF-Osservatorio Astronomico di Padova, Vicolo dell'Osservatorio 5, I-35122, Padova, Italy}
\affil{$^7$Institut d'Astrophysique de Paris, UMR7095, Universit{\'e} Pierre et Marie Curie, 98 bis Boulevard Arago, 75014 Paris, France}
\affil{$^8$Spitzer Science Center, California Institute of Technology 220-06, Pasadena, CA 91125, USA}
\affil{$^9$Institute for Astronomy, ETH Zurich, Wolfgang-Pauli-strasse 27, 8093 Zurich, Switzerland}
\affil{$^{10}$Dipartimento di Astronomia, Universit{\`a} di Bologna, Via Ranzani 1, 40127 Bologna, Italy}
\affil{$^{11}$Department of Astronomy, University of Massachusetts, Amherst, MA, USA}
\affil{$^{12}$Laboratoire d'Astrophysique de Marseille, BP 8, Traverse du Siphon, 13376 Marseille Cedex 12, France}
\affil{$^{13}$Center for Astrophysics, University of Science and Technology of China, Hefei, Anhui 230026, China}
\affil{$^{14}$Institute of Astronomy, University of Tokyo, Mitaka, Tokyo, Japan}
\affil{$^{15}$Department of Astronomy, Kyoto University, Kyoto 606-8502, Japan}
\affil{$^{16}$Institute for Astronomy, University of Hawaii, 2680 Woodlawn Drive, Honolulu, HI 96822, USA}
\affil{$^{17}$California Institute of Technology, MC 105-24, 1200 East California Boulevard, Pasadena, CA 91125, USA}
\affil{$^{18}$Subaru Telescope, National Astronomical Observatory of Japan, 650 North A'ohoku Place, Hilo, HI 96720, USA}
\affil{$^{19}$Research Center for Space and Cosmic Evolution, Ehime University, 2-5 Bunkyo-cho, Matsuyama 790-8577, Japan}

\email{masato.onodera@cea.fr}

\begin{abstract}
  We present observations of a very massive galaxy at $z=1.82$ which
  show that its morphology, size, velocity dispersion and stellar
  population properties that are fully consistent with those expected
  for passively evolving progenitors of today's giant ellipticals.
  These findings are based on a deep optical rest-frame spectrum
  obtained with the Multi-Object InfraRed Camera and Spectrograph
  (MOIRCS) on the Subaru telescope of a high-$z$ passive galaxy
  candidate (pBzK) from the COSMOS field, for which we accurately
  measure its redshift of $z=1.8230$ and obtain an upper limit on its
  velocity dispersion $\sigma_\star<326$~\kms{}.  By detailed stellar
  population modeling of both the galaxy broad-band SED and the
  rest-frame optical spectrum we derive a star-formation-weighted age
  and formation redshift of $t_\text{sf}\simeq1$--$2$~Gyr and
  $z_\text{form}\simeq2.5$--$4$, and a stellar mass of
  $M_\star\simeq3$--$4\times10^{11}\,M_\sun$. This is in agreement
  with a virial mass limit of $M_\text{vir}<7\times10^{11}\,M_\sun$,
  derived from the measured $\sigma_\star$ value and stellar
  half-light radius, as well as with the dynamical mass limit based on
  the Jeans equations.  In contrast with previously reported
  super-dense passive galaxies at $z\sim2$, the present galaxy at
  $z=1.82$ appears to have both size and velocity dispersion similar
  to early-type galaxies in the local Universe with similar stellar mass. 
  This suggests that $z\sim2$ massive and passive galaxies may exhibit 
  a wide range of properties, then possibly following quite different
  evolutionary histories from $z\sim2$ to $z=0$.
\end{abstract}


\section{Introduction}

Understanding the formation of massive elliptical galaxies remains a
crucial unsolved issue of galaxy evolution. The recent discovery and
the first redshift measurements, through deep ultraviolet (UV) rest-frame
spectroscopy, of a substantial population of passively evolving
galaxies at $z>1.4$
\citep[e.g.,][]{cimatti:2004,mccarthy:2004,daddi:2005:pbzk} have shown
that quenching of star formation in the most massive galaxies was
already well under way at $z\simeq2$.

A puzzling property of such objects has been revealed soon afterwards
with some of them being found to have a factor of $\simeq2-5$ smaller
effective radii compared to local early-type galaxies (ETGs) of the
same stellar mass
\citep[e.g.,][]{daddi:2005:pbzk,trujillo:2006,longhetti:2007,cimatti:2008,vandokkum:2008},
implying that they are $\gtrsim 10$ times denser than their possible
descendants in the local Universe.  Several alternative mechanisms
have been proposed to make such compact ETGs grow in size so to
finally meet the properties of local ETGs
\citep[e.g.,][]{khochfar:2006,fan:2008,naab:2009,labarbera:2009,feldmann:2010},
but no general consensus has yet emerged.

On the other hand, ETGs at $z>1.4$ with large effective radii, comparable to the local ETGs,
have also been found 
\citep[e.g.,][see also \citealt{daddi:2005:pbzk}]{mancini:2010,saracco:2009}, 
indicating a diversity of structural properties in the ETG population at $z\simeq2$. Moreover,
possible effects  have also been discussed  that could bias size estimates towards lower values
\citep[e.g.,][]{daddi:2005:pbzk,hopkins:2009,mancini:2010,pannella:2009}. 

An independent way to check these issues is by measuring stellar
velocity dispersions ($\sigma_\star$): if high-$z$ ETGs are really
super-dense, their $\sigma_\star$ should be much higher than that of local
ETGs of the same mass.  \citet{cappellari:2009:gmass} measured
$\sigma_\star$ from deep UV rest-frame spectroscopy of a sample of 9 ETGs at
$1.4<z<2.0$ from the GMASS survey \citep{cimatti:2008}, finding two
galaxies with similar stellar density and $\sigma_\star$ as local ETGs of
the same mass.  The remaining galaxies have higher stellar densities
and higher $\sigma_\star$ from their stacked spectrum, but still overlapping
with the densest local ETGs.

In this respect, near-infrared (NIR) spectroscopy offers a great
advantage for the galaxies at $z\gtrsim1.4$ as the optical break at
rest-frame $4000$~\AA{} and \cahk{} (the strongest spectral features
of passively evolving galaxies) are redshifted into NIR, and the
rest-frame optical continuum is much brighter than that in the
rest-frame UV.  Thus, \citet{vandokkum:2009} measured a velocity
dispersion of $\sigma_\star\simeq500$~\kms{} for a compact ETG at $z=2.186$
using a deep NIR spectrum from \citet{kriek:2009}. This value of
$\sigma_\star$ is much higher than that of the most massive local ETGs, and
would be consistent with the small half light radius measured for that
galaxy.

In this letter we present a rest-frame optical spectrum of a massive,
passively evolving high redshift galaxy candidate taken with the
Multi-Object InfraRed Camera and Spectrograph
\citep[MOIRCS;][]{ichikawa:2006:moircs,suzuki:2008:moircs} at Subaru
Telescope.  
A cosmology with $\Omega_{\text{M}}=0.3$, $\Omega_{\Lambda}=0.7$, and $H_0 = 70$
\kms~Mpc$^{-1}$ is assumed.

\section{Observations and Data Reductions}

We have obtained 4.7~hours of MOIRCS spectroscopy of 34 BzK-selected
galaxies \citep{daddi:2004:bzk} from the catalog of
\citet{mccracken:2010} in the COSMOS field. We used the zJ500 grism
with $0\farcs7$ slits, providing a resolution of $R\simeq500$ in the
$J$-band over the range 9500--16000\AA{}.  The primary aim of
our observation was to measure redshifts for passive BzK galaxies
(pBzKs) and locate them accurately in the COSMOS large scale
structure.  We preferentially selected the most massive pBzKs (which
are also the brightest in the NIR) to maximize the chance of
determining also physical information in addition to redshifts for
at least a fraction of them, in particular the galaxy (\obj)
discussed in this paper and already studied by \citet{mancini:2010}. 
The observations were made under partly cloudy conditions and with
$\sim1\farcs2$ seeing.  A sequence of 600s integrations were made in a
two position dithering pattern separated by $2\arcsec$.  The A0V-type
star HIP 55627 was observed to obtain (relative) flux calibration and
to correct for telluric absorption. The data were reduced using the
\textit{MCSMDP} pipeline (Yoshikawa et al., in preparation), including
flat-fielding by dome flat, sky subtraction between each exposure
pairs, bad pixel and cosmic ray rejection, distortion correction,
wavelength calibration (based on the OH telluric lines), residual sky subtraction
and finally co-addition with appropriate offsets and weights. The
2-dimensional spectra are flux-calibrated using the standard star
spectrum, and 1-dimensional spectra were extracted with the IRAF
\textit{apall} task using a $1\farcs9$ aperture.  The absolute flux
calibration was then obtained by normalizing to the $J$-band total
magnitude.

Whilst the results for the complete sample observed with MOIRCS will
be presented elsewhere we will concentrate here on the pBzK galaxy
\obj.  This galaxy one of the 12 ultra-massive high redshift ETG candidates in
\citet{mancini:2010}, with a photometric redshift of
$z_\text{phot}=1.71$ and very bright NIR magnitudes of
$J_\text{AB}=20.32$ and $K_\text{AB}=19.41$.  \citet{mancini:2010}
also report that the galaxy is non-detected at \textit{Spitzer}/MIPS
$24\,\micron$ to $80\,\mu\text{Jy}$ implying a star formation rate
$\text{SFR}\lesssim50\,M_\odot\text{yr}^{-1}$.  Using \textit{HST}/ACS
F814W imaging (rest-frame UV) they measure a S{\'e}rsic index $n=4.1$
and an effective radius of $\re=5.7$~kpc, consistent with the stellar
mass-size relation of local elliptical galaxies \citep{mancini:2010}.

Figure \ref{fig:specall} shows the resulting 1D and 2D MOIRCS spectra
of \obj.  The 4000~\AA{} break is clearly seen, together with strong
Balmer and metallic absorption lines, namely \cah+\heps, \cak, \hd,
\hg, \hzeta, G-band and CN+\htheta.  \hbsp falls in the region with
low atmospheric transmission and with strong OH-lines.  No emission
lines are observed.  While [\ion{O}{3}]$\lambda\lambda4959,5007$ falls
in a region with low atmospheric transmission,
[\ion{O}{2}]$\lambda3727$ is uncontaminated and its non-detection sets
a $3\sigma$ upper limit of SFR of $\simeq2.5\,M_\sun\text{yr}^{-1}$
(not corrected for extinction), using the \citet{kennicutt:1998}
conversion.  From the spectrum, the absorption line redshift is
measured as $z=1.8230\pm0.0006$.

\section{Results}

\subsection{Stellar Populations \label{sec:stellarpop}}

\begin{figure}
  \begin{center}
    \includegraphics[width=0.9\linewidth]{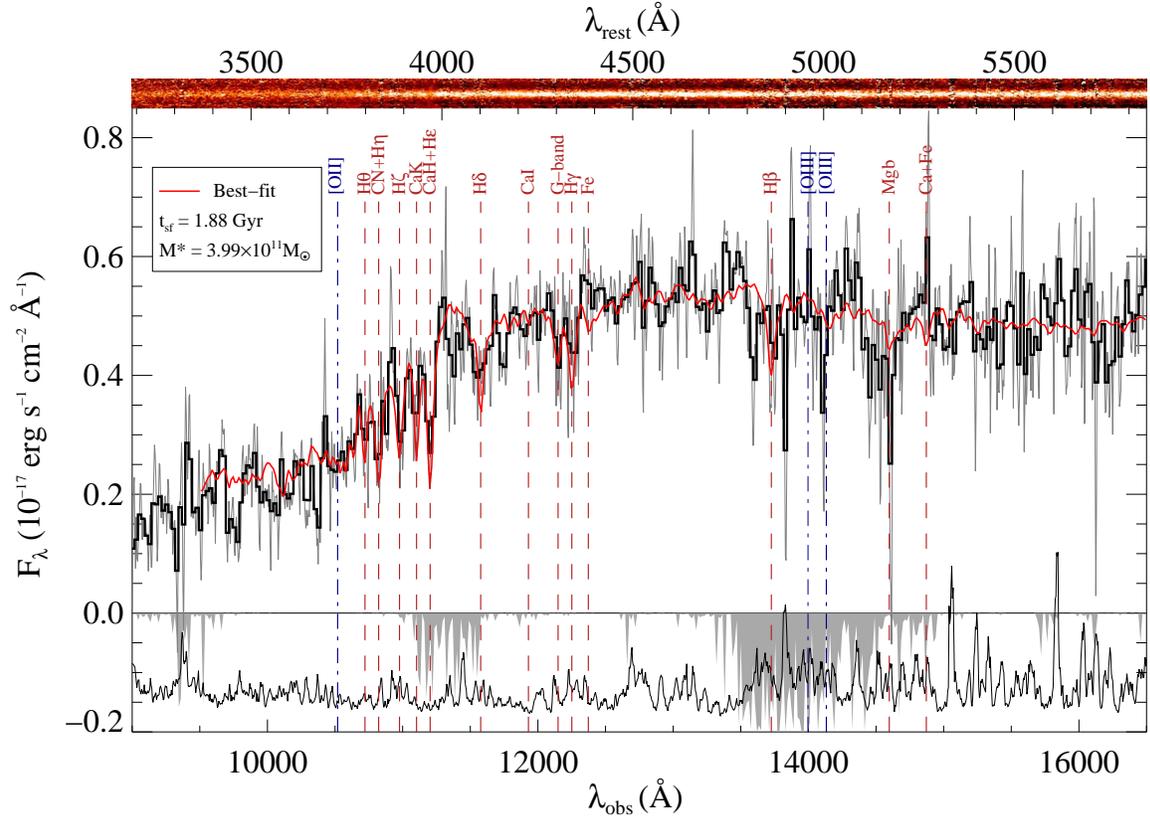}
  \end{center}
  \figcaption{
    MOIRCS  spectrum of \obj at $z_\text{spec}=1.82$. Top: $4''$ 2D spectrum; middle: 1D spectrum
    without smoothing (grey line) and with a 25\AA{} binning (black thick line); bottom: relative noise level (solid line) 
    and the sky transmission (shaded area).
    The red solid line shows the best-fit model (see \S\ref{sec:stellarpop}).
    Positions of major emission  and  absorption lines are indicated by dot-dashed (blue) 
    and dashed (red) lines, respectively, even when not detected. 
    \label{fig:specall}
  }
\end{figure}

Having determined the spectroscopic redshift, we proceeded to fit
stellar population templates, separately to the broad-band spectral energy distribution (SED) and to
the MOIRCS spectrum.  We allowed for a wide range of possible star
formation histories (SFHs), including: (1) instantaneous bursts, i.e.,
simple stellar populations (SSP); (2) constant SFRs for a duration
within 0.01--3.5~Gyr\,~\footnote{The age of the Universe at $z=1.82$
  is about 3.5~Gyr, given the adopted cosmology.}, terminated by SF
quenching and followed by passive evolution; (3) delayed,
exponentially declining SFH described as
$\text{SFR}(t,\tau)\propto(t/\tau^2)\exp(-t/\tau)$ with $\tau$ within
$0.01$--$2$~Gyr; (4) exponentially increasing SFH,
$\text{SFR}(t,\tau)\propto\exp(t/\tau)$ for a duration within
$t_q=0.1$--$3$~Gyr, followed by SF quenching and passive evolution. We
choose $\tau=0.72$~Gyr, corresponding to a stellar mass doubling time
of $\simeq 0.5$~Gyr, as suggested for $z\sim 2$ galaxies by the
existence of tight stellar mass-SFR relation, with
$\text{SFR}\propto\sim M_\star$ \citep{daddi:2007:sfr,renzini:2009}.
For all the SFHs, template ages were allowed to range in
$t=0.4$--$3.5$~Gyr.  We use a \citet{chabrier:2003} IMF.  
The fits were made with metallicities of 0.5$\times Z_\sun$, $Z_\sun$
and 2$\times Z_\sun$. To reduce the number of free
parameters we have assumed no dust extinction, appropriate for a
passively evolving galaxy, considering the strict upper limit on the
SFR that is set by the spectrum, an assumption that is validated by
the good fit that is achieved in the blue continuum (see Figure
\ref{fig:specall}).

The SED fitting was carried out for the broad-band $Biz$ data from
Subaru/Suprime-Cam \citep{capak:2007,taniguchi:2007}, $JHK$ data from
CFHT/WIRCAM \citep{mccracken:2010}, and the
\text{\textit{Spitzer}/IRAC} $3.6\micron$, $4.5\micron$ and
$5.8\micron$ data \citep{sanders:2007}.  Artificial errors of 0.05 mag
for the $BizJHK$ bands and 0.1 mag for the IRAC bands are added to the
observed errors in quadrature to account for systematics in zero-point
determinations, in the photometric measurements, and in the stellar population models. 
The templates for the
SED fitting are generated from population synthesis models by
\citet[][hereafter \citetalias{maraston:2005}]{maraston:2005}. 

The spectral resolution of the \citetalias{maraston:2005} models is
significantly lower than that of our MOIRCS spectrum. Hence, for
fitting the spectrum, we used templates from the \citet[][hereafter
\citetalias{bruzual:2003}]{bruzual:2003} spectral synthesis
library. Although these models might not account properly for TP-AGB
stars \citep[e.g.,][]{maraston:2006}, this effect is not significant at
4000\AA\ rest-frame, the wavelength range probed by our spectrum.  The
template spectra are Gaussian-broadened to an overall velocity
dispersion of 350~\kms{} (see \S\ref{sec:sigma}) to match that the
observed spectrum, and having fixed it the stellar population
parameters are derived with the $\chi^2$ over the observed wavelength
range 9500--16000\AA{}.

\begin{figure}
  \begin{center}
    \plotone{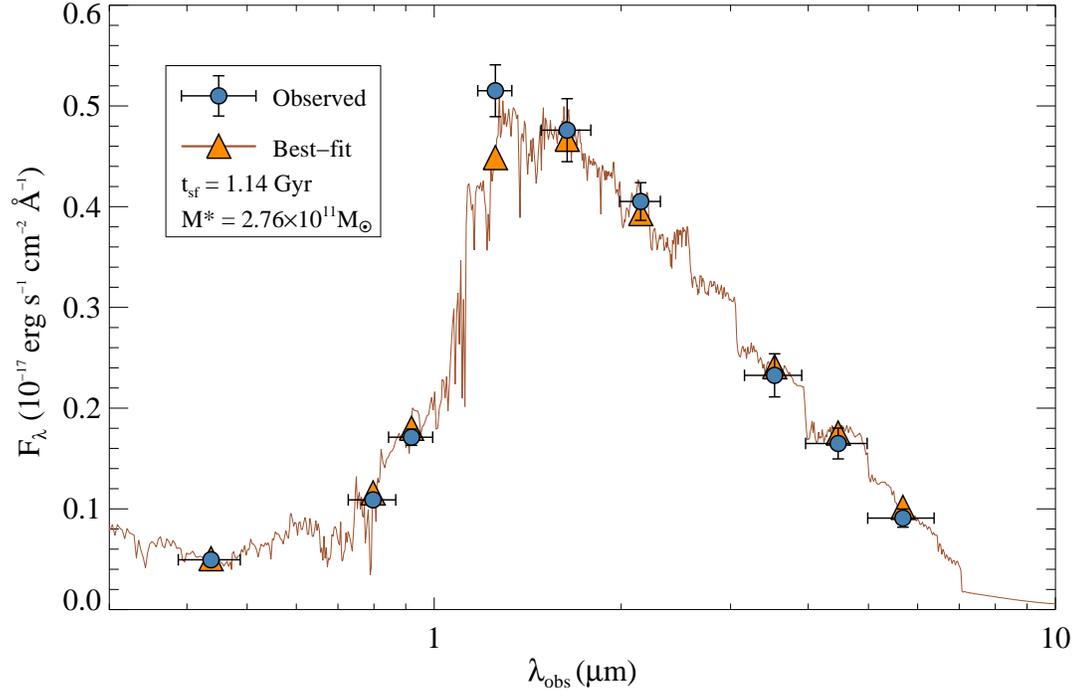}
  \end{center}
  \figcaption{
    Observed SED for \obj (blue circles with error bars) compared to the 
    best fit model from Maraston (2005; orange line and symbols).  
    The best fit was found for a SFH with delayed exponential star formation rate that continued for about 1.4~Gyr, 
    with a star-formation timescale of 0.13~Gyr, and for a solar metallicity. 
    The parameters of the best fit templates can be found in Table 1.  
    \label{fig:sedbest}
  }
\end{figure}

The stellar population parameters of the best-fit models from each
adopted SFH are listed in Table \ref{tab:fitting} and the best-fit
templates for the spectrum and SED are shown in Figures
\ref{fig:specall} and \ref{fig:sedbest}, respectively. The best fit
spectra have star-formation (SF) weighted ages of
$t_\text{sf}=1.14^{+0.73}_{-0.01}$~Gyr and stellar masses of
$M_\star=(2.76^{+0.82}_{-0.01})\times10^{11}\,M_\sun$ for the
broad-band SED, and $t_\text{sf}=1.88^{+0.01}_{-0.24}$~Gyr and stellar
mass of $M_\star=(3.99^{+0.10}_{-0.32})\times10^{11}\,M_\sun$ for the
spectrum.  The best fit results are from $Z=Z_\sun$ for the SED and
$Z=2\times Z_\sun$ for the spectrum. However, for the spectrum very
similar values are derived using solar metallicity models that result
in a slightly higher $\chi^2$ (Table~1). We notice that solar, or
slightly supersolar metallicities are appropriate for local elliptical
galaxies with stellar masses similar to galaxy \obj
\citep{thomas:2005}. The SF weighted age of $\simeq1$--$2$~Gyr
corresponds to an average formation redshift of $z_{\text{form}}\simeq2.5$--$4$, 
although the SF could have started much earlier.
In the case of the spectral fitting, all SFHs adopted here fit equally
well with $\chi^2\simeq1.3$ and $t_\text{sf}\simeq1$--$2$~Gyr,
consistent with the detection of strong Balmer absorption lines which
are most prominent in A-type stars.  Moreover, in the case of
the SED fits (apart for the SSP spectra) the various SFHs do not
give very different $\chi^2$ values and therefore it is not possible
to discriminate between them.  The same can be said for the derived
metallicities. If we allow for dust reddening, the best fitting
$t_\text{sf}$, $M_\star$ and $M/L_U$ would change by only $\lesssim10$\%{}, 
with some increase of the formal uncertainties within each class of SFH.

\subsection{Velocity Dispersion and Dynamical Modeling \label{sec:sigma}}

Our high S/N spectrum ($\simeq8.7$ per 60~\kms{} spectral interval in
the continuum) allows us to measure a velocity dispersion from the
absorption line width $\sigma_\text{obs}$, which is a combination of
the galaxy stellar velocity dispersion $\sigma_\star$ and the
instrumental resolution $\sigma_\text{instr}$. This S/N is comparable
to spectra of GMASS galaxies with successful individual $\sigma_\star$
determinations \citep{cappellari:2009:gmass}. Therefore we followed
the same approach of \citeauthor{cappellari:2009:gmass}, based on the
Penalized Pixel-Fitting method \citep[pPXF;][]{cappellari:2004:ppxf}.
The MILES stellar library containing 985 stars
\citep{sanchezblazquez:2006:miles} is adopted here since it provides
the best uniform and complete set of stars.

\begin{figure}
  \begin{center}
    \includegraphics[width=0.5\linewidth]{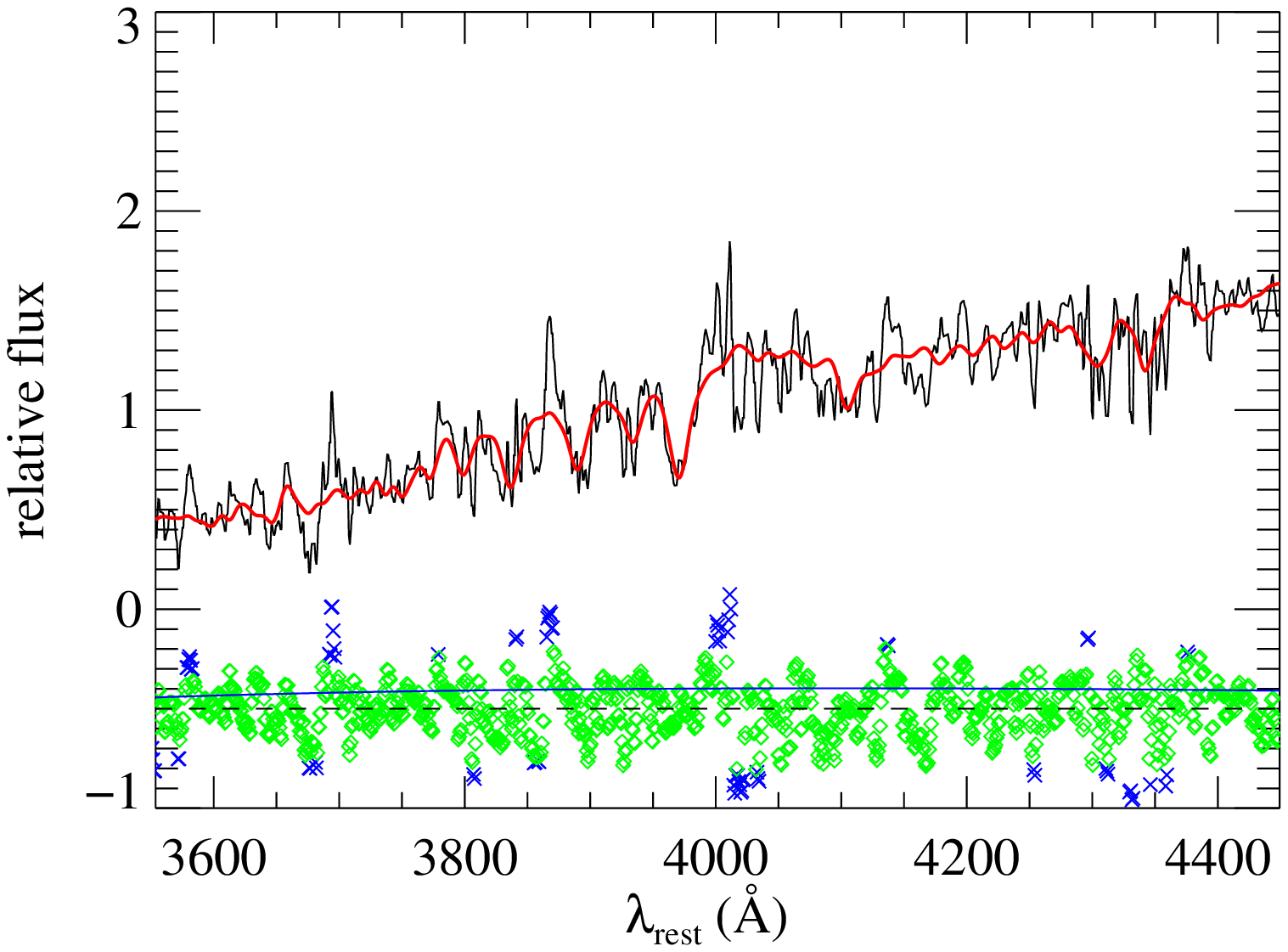}
    \includegraphics[width=0.5\linewidth]{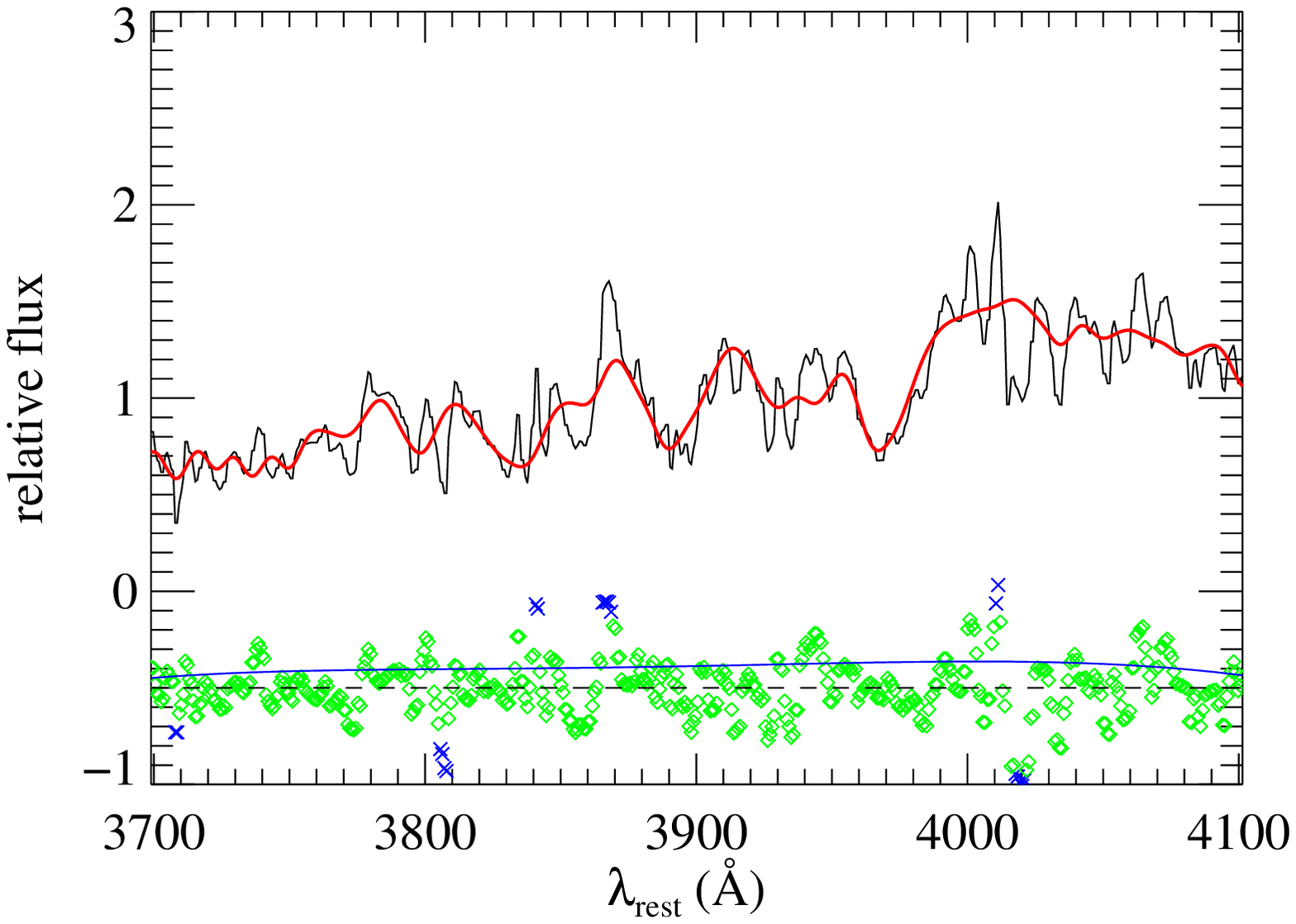}
  \end{center}
  \figcaption{
    Results of the pPXF fit for the stellar velocity dispersion of the galaxy. 
    The panels show the resulting fit for the full spectral range (top), and 
    for a wavelength range around \cahk{} (bottom). 
    The black solid line shows the observed spectrum, 
    the red solid line shows  the best-fit template, 
    and the green diamonds are the residuals (arbitrarily shifted).  
    The blue crosses indicate bad pixels rejected from the fitting. 
    The solid blue line indicates the estimated  $1\sigma$ noise level. 
    \label{fig:ppxf}
  }
\end{figure}

Figure \ref{fig:ppxf} shows the best-fit templates from pPXF, corresponding to 
$\sigma_{\text{obs}} = 350\pm30$ km s$^{-1}$ (random) $\pm$ 30 km s$^{-1}$ (systematic) 
for the rest-frame wavelength range of $3500-4450$ \AA. The random error ($1\sigma$ 
confidence) is determined as half of the interval in $\sigma_{\text{obs}}$ spanned by
68 out of 100 Monte Carlo realizations of the input spectrum. A rough estimate of
the systematic error is obtained as half of the interval in $\sigma_{\text{obs}}$
spanned by all repeated extractions of the kinematics using different, but equally
acceptable, combinations for the values of the degree (from 0--4) of the additive
and multiplicative polynomials in pPXF. Restricting the fit to the region with
the Balmer and \cahk lines ($3700-4100$ \AA\ in the rest-frame) gives 
$\sigma_{\text{obs}} = 300\pm50$ km s$^{-1}$ (total error), consistent with the
value derived from the full spectral range.

In order to derive $\sigma_\star$ we need to determine accurately the
instrumental resolution. To do this we have used our combined MOIRCS
spectrum without sky subtraction and simultaneously fitted Gaussian
profiles to a series of telluric OH-lines at
$\lambda\simeq11,500$~\AA{}, i.e., near the strongest absorption
features of the galaxy's spectrum. The central wavelength for each
OH-line was taken from \citet{rousselot:2000} and we left
$\sigma_\text{instr}$, OH-line intensities, and constant baseline as
free parameters.  The fitting procedure reproduces the observed sky
spectra very well.  Over $\lambda = 3500$--$4450$~\AA{}, the
instrumental resolution changes from 270 \kms{} to 330 \kms{}.  We
adopt $\sigma_\text{instr}=300 \pm 7$ \kms{} (random) $\pm 30$ \kms{}
(systematic).

The derived galaxy stellar velocity dispersion is
$\sigma_\star=\sqrt{\sigma_{\text{obs}}^2-\sigma_{\text{instr}}^2}$, which
gives $\sigma_\star=180\pm59$ \kms{} (random) $\pm 87$ \kms{}
(systematic).  The relatively large uncertainties in $\sigma_{\text{obs}}$, 
and as $\sigma_{\text{obs}}$ is close to $\sigma_{\text{instr}}$,
we cannot place a lower limit to $\sigma_\star$.  However, we can
derive a $1\sigma$ upper limit of $\sigma_\star<326$ \kms{} (or
$\sigma_\star<385$ \kms{} at the $2\sigma$ level), which is consistent
with both determinations.  If Balmer lines suffer from fill-in from
emission lines $\sigma_\star$ could be somewhat lower.

From the stellar velocity dispersion the virial mass can be calculated
as $M_\text{vir}=C \re \sigma_\star^2/G$.  We have set $C=5$ as
empirically calibrated on local galaxies with state-of-the-art
dynamical modeling \citep{cappellari:2006}, with the velocity
dispersion being measured within a large aperture ($\sim 1\re$) as in
our case.  The effective radius $\re$ was measured by
\citet{mancini:2010} from the \hst/ACS F814W image ($\simeq
2900$~\AA{} in the rest-frame) as $0.68'' \pm 0.07''$ or $5.7 \pm
0.6$~kpc at $z=1.82$.  Thus the upper limit of the virial mass is
derived as $M_\text{vir}<7.0\times10^{11}\,M_\sun$.

We have also constructed a dynamical model based on axisymmetric Jeans
dynamical models as those used to model the GMASS galaxies by
\citet{cappellari:2009:gmass}, adopting a multi-Gaussian expansion
\citep[MGE;][]{emsellem:1994}.  This method has the advantage that the
derived $M/L$ is virtually insensitive to possible underestimation of
the size, which can be a possibility at high redshifts Considering the
bolometric surface brightness dimming of $(1+z)^4$, a factor $(1+z)$
coming from the source redshifting and the $K$-correction between
rest-frame 2900~\AA{} to rest-frame $U$-band, we derived a rest-frame
$U$-band luminosity of $L_U=6.7\times10^{11}\,L_\sun$.  The second
moment of the velocity $V^{2}_{\text{rms}}=V^2+\sigma_\star^2$ was also
estimated (assuming $\beta_z=0$ and axisymmetry) by using the Jeans
anisotropic MGE (JAM) method \citep{cappellari:2008}.  The upper limit
for the dynamical $M/L_U$ can be calculated by
$(M/L_U)_\text{Jeans}=(\sigma_\star/V_\text{rms})^2<1.0$, which can be
converted into the upper limit of the dynamical mass from the JAM
model as $M_\text{Jeans}=L_U \times (M/L_U)_\text{Jeans}<6.8 \times
10^{11}\,M_\odot$.  Therefore the virial mass and Jeans mass agree
well though both of them are upper limits.  A JAM model constructed
from a noiseless model with the best fitting S{\'e}rsic parameters of
\citet{mancini:2010}, as opposed to the actual ACS image, gives the
same $(M/L_U)_\text{Jeans}$ within 1\%. This is due to the robustness
of the central $M/L$ recovered using dynamical models (in contrast to
virial estimates) to photometric uncertainties at large radii
\citep[e.g., \S{}3.2 in][]{cappellari:2009:gmass}.

\section{Discussion and Conclusions}

\begin{figure}
  \begin{center}
    \plotone{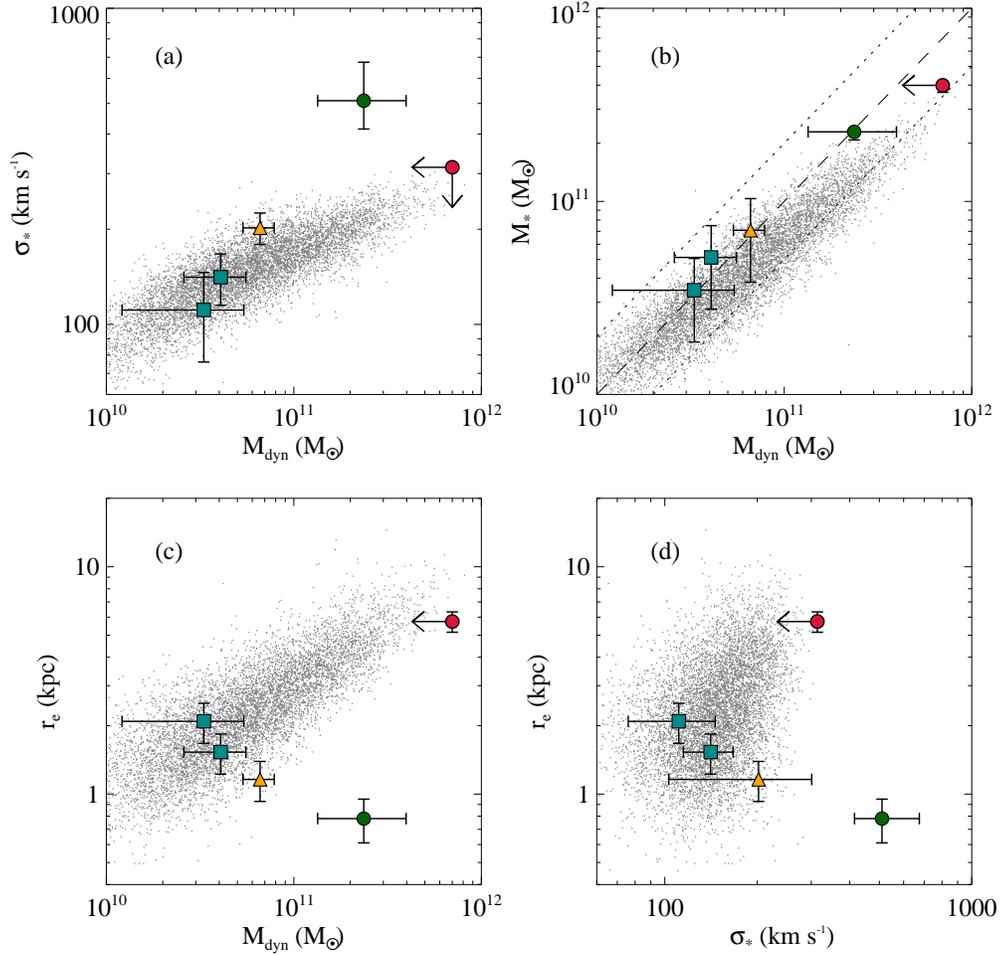}
  \end{center}
  \figcaption{
    A comparison of the properties of high-redshift ETGs for which
    the velocity dispersion has been measured so far (symbols with error bars) 
    with those of elliptical galaxies selected from SDSS at $z\simeq0.06$ (gray dots).  
    (\textit{a}): the stellar velocity dispersions vs.\ the  virial masses. 
    (\textit{b}): comparison between virial and stellar masses; 
    the diagonal dashed line corresponds to equality of the two masses, 
    and the dotted lines show a range by a factor of 2. 
    (\textit{c}): the effective radii vs.\ the virial masses. 
    (\textit{d}): the effective radii vs.\ stellar velocity dispersions. 
    In all panels, the red filled circle represents the galaxy studied here,  the
    green filled circle shows the object studied by \citet{vandokkum:2009}, 
    the blue squares represent the two GMASS galaxies with individual $\sigma_\star$ measurements, 
    and the yellow triangle represents the properties from the stacked GMASS spectrum 
    \citep[taken from][]{cappellari:2009:gmass}.
   \label{fig:dynprop}
  }
\end{figure}

Figure \ref{fig:dynprop} compares the properties of the galaxy \obj
and other $z\simeq2$ galaxies for which the same quantities have been
measured \citep{cappellari:2009:gmass,vandokkum:2009}.  The figure
includes ETGs at $z\simeq0.06$, selected from the Sloan Digital Sky
Survey (SDSS) on the base of their red $u-g$ color and high S{\'e}rsic
index $n\simeq4$ \citep{blanton:2005}.  The dynamical and stellar
masses agree very well for the high-$z$ objects, within a factor of
$\lesssim 2$.  Note that our massive galaxy has physical properties in
good agreement with those of local ETGs of similar stellar mass. Our
galaxy provides a second example of a very massive passively evolving
system for which a stellar velocity dispersion has been measured (the
GMASS objects of \citeauthor{cappellari:2009:gmass}, which in Figure
\ref{fig:dynprop} also lie on the $z=0$ scaling relations, but have
stellar masses below $10^{11}\,M_{\sun}$).  The ``normal'' size and
velocity dispersion of our massive ETG is strikingly in contrast with
the extreme properties (i.e., a very high $\sigma_\star =
510_{-95}^{+163}$~\kms{} and a small $\re=0.78\pm0.17$~kpc) of the
galaxy studied by \citet{vandokkum:2009} with similar stellar mass
($2\times10^{11}\,M_\sun$).  This suggests a substantial diversity in
the physical properties of the \textit{massive} ETG population at
$z\sim2$ including ``immature'', albeit virialized, systems --- which
will have to evolve into normal $z=0$ massive galaxies through some
physical processes which decrease their velocity dispersion and
increase their sizes --- as well as ``mature'' ETGs, already on the
scaling relationships of $z=0$ ETGs.  It is clear that many
more observations of similar galaxies are required to establish
which kind of ETG is commonest at high redshift: either the
compact/high-$\sigma_\star$ objects like those found by
\citet{vandokkum:2009}, or the apparently normal, low-$\sigma_\star$
objects presented in this paper.  Also, nothing prevents our
particular object to evolve further from its present state which
mimics that of local ellipticals of the same mass.  For example, it
may grow further and become a \textit{brightest cluster galaxy}
(BCG) or a cD galaxy.  For this reason, it would be important to
estimate the volume number density of similar objects at high
redshifts, and compare it to that of BCGs and cD galaxies.

To conclude, both very compact ETGs and ETGs following the local
stellar mass-size and stellar mass-$\sigma_\star$ relations appear to
co-exist at $z>1.4$ \citep[see also ][]{mancini:2010}.  However the
number of high-$z$ ETGs with individual measurement of the velocity
dispersion is still extremely small. Increasing their sample is of
great importance to understand the evolution of these galaxies, and
in particular how and when they acquire their final structural and
dynamical configuration.  This paper demonstrates that with reasonable
telescope time several absorption features can be detected in the
rest-frame optical spectrum of the high-$z$ ETGs, from which (at least
for the most massive ETGs) the velocity dispersion and several stellar
population properties can be derived.

\acknowledgments
We are grateful to Tomohiro Yoshikawa for providing MCSMDP 
before publication. We thank 
Subaru telescope staff for help with our observations. 
We acknowledge funding ERC-StG-UPGAL-240039, ANR-07-BLAN-0228, ANR-08-JCJC-0008
and a Grant-in-Aid for 
Science Research (No. 19540245) by the Japanese Ministry 
of Education, Culture, Sports, Science and Technology. 
AR is grateful to the Institute of Astronomy of 
ETH Z\"urich for its kind hospitality. 
MC acknowledges support from a STFC Advanced Fellowship (PP/D005574/1). 


\clearpage


\clearpage
\begin{landscape}
\begin{deluxetable}{lccccccccccccc}
  \tabletypesize{\scriptsize}
  \tablewidth{0pt}
  \tablecolumns{14}
  \tablecaption{Best-fit Stellar Population Parameters for the SED and the Spectrum\label{tab:fitting}}
  \tablehead{
    \colhead{} & \multicolumn{6}{c}{SED} & \colhead{} & \multicolumn{6}{c}{Spectrum} \\
    \cline{2-7} \cline{9-14} 
    \colhead{SFH} &
    \colhead{$T_\text{onset}$} &
    \colhead{$\tau$ or $t_q$} &
    \colhead{$t_\text{sf}$} &
    \colhead{$M_\star$} &
    \colhead{$M/L_U$} &
    \colhead{$\chi^2$} &
    \colhead{} &
    \colhead{$T_\text{onset}$} &
    \colhead{$\tau$ or $t_q$} &
    \colhead{$t_\text{sf}$} &
    \colhead{$M_\star$} &
    \colhead{$M/L_U$} &
    \colhead{$\chi^2$} 
    \\
    \colhead{} &
    \colhead{(Gyr)} &
    \colhead{(Gyr)} &
    \colhead{(Gyr)} &
    \colhead{($10^{11}M_\sun$)} &
    \colhead{($(M/L_U)_\sun$)} &
    \colhead{} &
    \colhead{} &
    \colhead{(Gyr)} &
    \colhead{(Gyr)} &
    \colhead{(Gyr)} &
    \colhead{($10^{11}M_\sun$)} &
    \colhead{($(M/L_U)_\sun$)} &
    \colhead{}
    \\
    \colhead{(1)} &
    \colhead{(2)} &
    \colhead{(3)} &
    \colhead{(4)} &
    \colhead{(5)} &
    \colhead{(6)} &
    \colhead{(7)} &
    \colhead{} &
    \colhead{(8)} &
    \colhead{(9)} &
    \colhead{(10)} &
    \colhead{(11)} &
    \colhead{(12)} &
    \colhead{(13)}
  }
  \startdata
  \multicolumn{14}{c}{$Z=0.5Z_\sun$} \\
  \cline{1-14}
  All                 & \nodata & \nodata& $2.54^{+0.24}_{-0.16}$ & $4.84^{+0.58}_{-0.30}$ & $1.11^{+0.11}_{-0.07}$ & $\phn2.35$ & 
                      & \nodata & \nodata& $1.16^{+0.86}_{-0.01}$ & $2.33^{+1.10}_{-0.02}$ & $0.45^{+0.21}_{-0.01}$ & $1.34$ \\
  SSP                 &  $0.7$  & \nodata& $0.70^{+0.01}_{-0.01}$ & $1.58^{+0.01}_{-0.01}$ & $0.35^{+0.01}_{-0.01}$ & $\phn9.25$ & 
                      &  $1.2$  & \nodata& $1.20^{+0.01}_{-0.03}$ & $2.40^{+0.01}_{-0.01}$ & $0.47^{+0.01}_{-0.01}$ & $1.35$ \\
  const. + quenching  &  $3.4$  & $3.00$ & $1.90^{+0.03}_{-0.61}$ & $2.83^{+0.16}_{-0.67}$ & $0.64^{+0.02}_{-0.15}$ & $\phn3.99$ & 
                      &  $1.2$  & $0.09$ & $1.16^{+0.86}_{-0.04}$ & $2.33^{+1.24}_{-0.03}$ & $0.45^{+0.23}_{-0.01}$ & $1.34$ \\
  delayed exponential &  $3.2$  & $0.33$ & $2.54^{+0.24}_{-0.56}$ & $4.84^{+0.58}_{-1.20}$ & $1.11^{+0.11}_{-0.24}$ & $\phn2.35$ & 
                      &  $1.2$  & $0.02$ & $1.16^{+0.13}_{-0.04}$ & $2.32^{+0.25}_{-0.04}$ & $0.45^{+0.04}_{-0.01}$ & $1.34$ \\
  exp. + quenching    &  $3.4$  & $2.87$ & $1.20^{+0.03}_{-0.36}$ & $1.98^{+0.10}_{-0.33}$ & $0.45^{+0.02}_{-0.08}$ & $\phn6.84$ & 
                      &  $1.2$  & $0.10$ & $1.15^{+0.39}_{-0.02}$ & $2.32^{+0.61}_{-0.03}$ & $0.45^{+0.11}_{-0.01}$ & $1.34$ \\
  \cline{1-14}
  \multicolumn{14}{c}{$Z=Z_\sun$} \\
  \cline{1-14}
  All                 & \nodata & \nodata& $1.14^{+0.73}_{-0.01}$ & $2.76^{+0.82}_{-0.01}$ & $0.63^{+0.21}_{-0.01}$ & $\phn2.20$ & 
                      & \nodata & \nodata& $1.95^{+0.01}_{-0.09}$ & $3.91^{+0.07}_{-0.16}$ & $0.77^{+0.02}_{-0.04}$ & $1.31$ \\
  SSP                 &  $0.6$  & \nodata& $0.60^{+0.01}_{-0.01}$ & $1.62^{+0.01}_{-0.01}$ & $0.34^{+0.01}_{-0.01}$ & $16.22$ & 
                      &  $1.0$  & \nodata& $1.00^{+0.01}_{-0.01}$ & $2.75^{+0.02}_{-0.03}$ & $0.54^{+0.01}_{-0.01}$ & $1.33$ \\
  const. + quenching  &  $3.4$  & $3.05$ & $1.88^{+0.02}_{-0.55}$ & $3.36^{+0.11}_{-0.75}$ & $0.75^{+0.02}_{-0.17}$ & $\phn2.28$ & 
                      &  $3.5$  & $3.11$ & $1.95^{+0.02}_{-0.92}$ & $3.91^{+0.16}_{-1.19}$ & $0.77^{+0.04}_{-0.23}$ & $1.31$ \\
  delayed exponential &  $1.4$  & $0.13$ & $1.14^{+0.70}_{-0.08}$ & $2.76^{+1.52}_{-0.24}$ & $0.63^{+0.40}_{-0.05}$ & $\phn2.20$ & 
                      &  $1.2$  & $0.07$ & $1.06^{+0.05}_{-0.04}$ & $2.78^{+0.12}_{-0.14}$ & $0.54^{+0.02}_{-0.03}$ & $1.32$ \\
  exp. + quenching    &  $1.7$  & $1.15$ & $0.98^{+0.17}_{-0.06}$ & $2.19^{+0.16}_{-0.16}$ & $0.48^{+0.03}_{-0.03}$ & $\phn6.62$ & 
                      &  $1.2$  & $0.31$ & $1.03^{+0.38}_{-0.04}$ & $2.74^{+0.62}_{-0.11}$ & $0.54^{+0.12}_{-0.02}$ & $1.32$ \\
  \cline{1-14}
  \multicolumn{14}{c}{$Z=2Z_\sun$} \\
  \cline{1-14}
  All                 & \nodata & \nodata& $0.97^{+0.86}_{-0.27}$ & $2.22^{+1.29}_{-0.26}$ & $0.52^{+0.34}_{-0.07}$ & $\phn4.32$ & 
                      & \nodata & \nodata& $1.88^{+0.01}_{-0.24}$ & $3.99^{+0.10}_{-0.32}$ & $0.80^{+0.03}_{-0.07}$ & $1.29$ \\
  SSP                 &  $0.4$  & \nodata& $0.40^{+0.01}_{-0.01}$ & $1.43^{+0.01}_{-0.01}$ & $0.31^{+0.01}_{-0.01}$ & $34.92$ &
                      &  $0.8$  & \nodata& $0.80^{+0.02}_{-0.01}$ & $2.65^{+0.09}_{-0.04}$ & $0.53^{+0.02}_{-0.01}$ & $1.31$ \\
  const. + quenching  &  $1.8$  & $1.62$ & $0.99^{+0.85}_{-0.45}$ & $2.35^{+1.25}_{-0.70}$ & $0.55^{+0.32}_{-0.18}$ & $\phn4.46$ & 
                      &  $3.5$  & $3.25$ & $1.88^{+0.01}_{-0.83}$ & $3.99^{+0.16}_{-1.07}$ & $0.80^{+0.04}_{-0.22}$ & $1.29$ \\
  delayed exponential &  $1.0$  & $0.10$ & $0.80^{+0.01}_{-0.01}$ & $2.39^{+0.01}_{-0.01}$ & $0.61^{+0.01}_{-0.01}$ & $\phn9.39$ & 
                      &  $1.1$  & $0.10$ & $0.90^{+0.03}_{-0.09}$ & $2.78^{+0.12}_{-0.17}$ & $0.55^{+0.03}_{-0.03}$ & $1.31$ \\
  exp. + quenching    &  $2.9$  & $2.58$ & $0.97^{+0.05}_{-0.32}$ & $2.22^{+0.17}_{-0.46}$ & $0.52^{+0.04}_{-0.12}$ & $\phn4.32$ & 
                      &  $3.3$  & $3.01$ & $1.80^{+0.02}_{-0.81}$ & $3.95^{+0.15}_{-1.11}$ & $0.79^{+0.04}_{-0.22}$ & $1.29$ 
  \enddata
  \tablecomments{
    Column 1: star-formation history (see \S\ref{sec:stellarpop}); 
    Column 2,8: elapsed time since the onset of star-formation; 
    Column 3,9: star-formation timescale in the
    case of delayed exponential SFH and quenching time for SFHs with constant SFR+quenching and 
    exponentially increasing SFR+quenching;
    Column 4,10: star-formation weighted age defined by $\int_0^T (T-t) \phi(t)\;dt / \int_0^T \phi(t)\;dt$ where 
    $T$ is $T_\text{onset}$ and $\phi(t)$ is SFR; 
    Column 5,11: stellar mass; 
    Column 6,12: rest-frame $U$-band mass-to-light ratio;
    Column 7,13: reduced $\chi^2$ for the best-fit template. 
    }
\end{deluxetable}
\clearpage
\end{landscape}

\end{document}